\documentclass[12pt]{article}

\title{Hirota Bilinear Method and Relativistic Dissipative Soliton Solutions in Nonlinear Spinor Equations}

\author{Oktay K Pashaev \\
Department of Mathematics \\ Izmir Institute of Technology \\ Izmir 35430, Turkey}

\begin{document}

\maketitle              

\begin{abstract}
 A new relativistic integrable nonlinear model for real, Majorana type spinor fields in 1+1 dimensions, gauge equivalent to Papanicolau spin model, defined on the one sheet hyperboloid is introduced. By using the double numbers, the model is represented as hyperbolic complex  valued relativistic massive Thirring type model. By Hirota's bilinear method, an exact
one and two dissipative  soliton solutions of this model are constructed. 
Calculation of first three integrals of motion for one dissipaton solution shows that the last one  represents a particle-like nonlinear excitation, with relativistic dispersion and highly nonlinear mass. A nontrivial solution of the system of algebraic equations, showing fusion and fission of relativistic dissipatons is found. Asymptotic analysis of exact two dissipaton solution confirms resonant character of our dissipaton interactions.
\end{abstract}



 %
\section{Introduction}

For description of black holes in low dimensional Jackiw-Teitelboim gravity model, two different, but equivalent approaches were proposed. One approach is based on the Resonant NLS equation (RNLS), the Nonlinear Schr\"odinger equation with quantum potential \cite{PL2002}, wich admits solutions in the form of  envelope soliton resonances. In another one, we have the system of real valued reaction-diffusion equations \cite{MPS1997}, \cite{MPS1998-1}, gauge equivalent to
the Heisenberg model for spin variable ${\bf s}$, belonging to  the one sheet hyperboloid $SL(2,R)/O(1,1)$. This system is integrable system, with specific type of solutions called the dissipatons \cite{Pashaev1997}. Dissipatons represent non-relativistic particles, with resonant interaction, allowing fusion and fission of single dissipatons and creating a reach, the web type structure of the interaction. In present paper, by using another type of spin model on the same hyperboloid, we construct new model of dissipatons, representing relativistic particles with resonant interaction.

\section{The Real Spinor Model}

\subsection{SL(2,R) spin model}
We start from 
nonlinear spin model corresponding to time independent Landau-Lifshitz equation in moving frame \cite{MPS1994}, \cite{MPS1998-2},
\begin{equation}
v^\mu \partial_\mu {\bf s} = {\bf s} \wedge \partial^\mu \partial_\mu {\bf s},\label{sigmamodel}
\end{equation}
for spin vector field ${\bf s}(x^0, x^1) \in SL(2,R)/O(1,1)$, where  the velocity vector ${\bf v} = (v^0, v^1)$  is constant  and  the space-time psudo-Euclidean metric is $diag(1,-1)$. The model  represents 
non-compact one sheet hyperbolic version of compact Papanicolau spin model \cite{Papa} . In terms of the light cone variables $v^+ = \frac{1}{2}(v^0 + v^1)$, $v^- = \frac{1}{2}(v^0 - v^1)$,
we have equation
\begin{equation}v^{+}\partial_{+}{\bf s} +
  v^{-}\partial_{-}{\bf s}
= {\bf s}\wedge \partial_{+}\partial_{-}{\bf s}, \label{sigmamodel2}\end{equation}
where $\partial_{\pm} = \partial_{0} \pm \partial_{1} $.

\subsection{Integrable real spinor model}

In present work we treat  the case of constant
"time-like"  velocity two-vector with length $\mu_0$,
\begin{equation}
v^{+} v^{-} = \frac{1}{4} [(v^0)^2 - (v^1)^2] \equiv \mu^2_0 >0. \label{velocity}
\end{equation}

The model (\ref{sigmamodel2}) with constant velocity two-vector ${\bf v} = (v^0, v^1)$, satisfying (\ref{velocity}), is gauge equivalent to
the real valued analog of Thirring model
 \begin{eqnarray}
- \partial_- p^+ + \mu_0\, q^+ + \frac{1}{\mu_0} q^+ q^- p^+ &=& 0, \label{1p}\\
 \partial_- p^- + \mu_0\, q^- +  \frac{1}{\mu_0} q^+ q^- p^-& = &0, \label{2p}\\
- \partial_+ q^+ + \mu_0\, p^+ + \frac{1}{\mu_0} p^+ p^- q^+ &= &0, \label{1q}\\
 \partial_+ q^- + \mu_0 \,p^- + \frac{1}{\mu_0} p^+ p^- q^-& =& 0. \label{2q}
\end{eqnarray}

 The system of equations (\ref{1p})-(\ref{2q}) is equivalent to
the compatibility conditions
$$
\partial_+ J_- -\partial_- J_+ + [J_+, J_-] =0,
$$
 for the linear system of equations
$$
\partial_{\pm} \Phi = \Phi J_{\pm},
$$
where 
the Lax pair in the zero-curvature condition form is
\begin{eqnarray}
J_{+} &=& \frac{1}{2}\left(-\lambda^{2} + \frac{1}{\mu_0}
p^+ p^-\right)\sigma_{3} + \lambda
\left(\begin{array}{cc}
0& \frac{1}{\mu_0}p^- \\ 
p^+&0 
\end{array}\right), \label{J1}\\
J_{-} &=& \frac{1}{2}\left(- \frac{\mu^2_0}{\lambda^{2}} + \frac{1}{\mu_0}q^+ q^-\right)\sigma_{3} + \frac{1}{\lambda}
\left(\begin{array}{cc} 
0& q^- \\ 
\mu_0 q^+&0
 \end{array}\right), \label{J2}
\end{eqnarray}
and $\lambda$ is the spectral parameter.

\section{Hyperbolic Complex Thirring model}

By expanding $ q^\pm = u_1 \pm v_1,\,\,\,\,p^\pm = u_2 \pm v_2$, we combine new functions to the pair of hyperbolic complex valued functions (known also as the double number valued functions) \cite{Yaglom}, 
\begin{equation}
\chi_1 = u_1 + j v_1,\,\,\,\,\chi_2 = u_2 + j v_2,
\end{equation}
where the hyperbolic imaginary unit satisfies
\begin{equation}
j^2 =1, \,\,\,\,\bar j = - j.
\end{equation}
The corresponding conjugate functions are
\begin{equation}
\bar \chi_1 = u_1 - j v_1,\,\,\,\,\bar\chi_2 = u_2 - j v_2,
\end{equation}
so that
\begin{equation}
\bar \chi_1 \chi_1 = |\chi_1|^2 = u^2_1 - v^2_1,\,\,\,\,\bar \chi_2 \chi_2 = |\chi_2|^2 = u^2_2 - v^2_2.
\end{equation}

The system of four equations (\ref{1p})-(\ref{2q}) can be represented in the form of two hyperbolic complex equations
\begin{eqnarray}
- j\partial_+ \chi_1 + \mu_0 \chi_2  + \frac{1}{\mu_0}|\chi_2|^2  \chi_1  &=& 0, \label{hT1}\\
 -j \partial_- \chi_2 + \mu_0 \chi_1  + \frac{1}{\mu_0}|\chi_1|^2  \chi_2  &=& 0. \label{hT2}
\end{eqnarray}

The system (\ref{hT1})-(\ref{hT2}) is equivalent to hyperbolic complex Thirring model
\begin{equation}
(-j \gamma^\nu \partial_\nu + m) \Psi + g^2 \gamma^\nu (\bar\Psi \gamma_\nu \Psi) \Psi =0,  \label{HT}
\end{equation}
for hyperbolic complex spinor field $\Psi$ and corresponding hyperbolic Dirac conjugate of it $\bar\Psi$,
\begin{equation}
\Psi = \left( \begin{array}{c} \psi_1 \\  \psi_2 \end{array}     \right),\,\,\,\,
\bar\Psi = \Psi^+ \gamma^0 = \left( \begin{array}{cc} \bar\psi_1 &  \bar\psi_2 \end{array}     \right) \gamma^0 = \left( \begin{array}{cc} \bar\psi_2 &  \bar\psi_1 \end{array}     \right),
\end{equation}
where we denote $\chi_1 \equiv g \sqrt{2 m}\psi_1$, $\chi_2 \equiv g \sqrt{2 m}\psi_2$, $\mu_0 \equiv m$.
Here Dirac's matrices are defined as
\begin{equation}
\gamma^0 = \left(\begin{array}{cc} 
0& 1\\ 
1&0
 \end{array}\right),\,\,\,
\gamma^1 = \left(\begin{array}{cr} 
0& -1\\ 
1&0
 \end{array}\right),
\end{equation}
the Minkowski space-time metric is $diag (1, -1)$ and 
\begin{equation}
\gamma^0 \partial_0 + \gamma^1 \partial_1 = \left(\begin{array}{cc} 
0& \partial_-\\ 
\partial_+&0
 \end{array}\right).
\end{equation}

As would be shown below, this hyperbolic complex counterpart to usual elliptic complex Thirring model,  in contrast to the last one admits resonant interaction of dissipative solitons. We notice also that the hyperbolic Thirring model (\ref{HT}) allows the Lax pair in terms of $2\times 2$ hyperbolic complex functions, while the equivalent real system (\ref{1p})-(\ref{2q}) has $2\times 2$
Lax pair (\ref{J1}),(\ref{J2}), 
but for the real valued functions.

\section{Hirota Bilinear Form }

By introducing six real functions $g^\pm, h^\pm, f^\pm$, we represent our field variables as
  $$p^\pm = \frac{g^\pm}{f^\mp} = \frac{g^\pm f^\pm}{f^\pm f^\mp},\,\,\,\,q^\pm = \frac{h^\pm}{f^\pm} = \frac{h^\pm f^\mp}{f^\mp f^\pm}.$$
Substituting to the system (\ref{1p})-(\ref{2q}) and using Hirota's derivatives
$$
D^k_t D^l_x (a (x, t)\cdot b(x, t)) = (\partial_t - \partial_{t'})^k (\partial_x - \partial_{x'})^l a(x, t) b(x', t')|_{x=x', t=t'},
$$
we split equations in form of the bilinear system.

The system of equations (\ref{1p})-(\ref{2q}) can be represented as bilinear system
\begin{eqnarray}
\pm D_t (g^\pm \cdot f^\pm) + \mu_0 \,h^\pm f^\mp &=& 0, \label{b1}\\
\mp D_x (h^\pm \cdot f^\mp) + \mu_0 \,g^\pm f^\pm &=& 0, \label{b2}\\
\mu_0 D_t (f^+ \cdot f^-) +  h^+ h^- &=& 0,\label{b3} \\
 \mu_0 D_x (f^+ \cdot f^-) + g^+ g^- &=& 0.\label{b4} 
\end{eqnarray}

Here $x^0 \equiv T$, $x^1 \equiv X$ are time and space coordinates in laboratory coordinate systems, 
$
x = \frac{1}{2}(X+T),$ $t = \frac{1}{2} (X-T)$ 
are the light-cone coordinates, so that $\partial_+ = \partial_x, \partial_- = - \partial_t$.

For calculating field densities, from Eqs. (\ref{b3}), (\ref{b4})  follow formulas
$$q^+ q^-(x,t) = - \mu_0 \left( \ln \frac{f^+}{f^-}\right)_t, \,\,\,\, p^+ p^-(x,t) = - \mu_0\left( \ln \frac{f^+}{f^-}\right)_x .$$

\section{The One Dissipaton Solution}

 By Hirota expansion in parameter $\epsilon$,
\begin{eqnarray*}
g^\pm(x,t)& =& \epsilon g_1^\pm(x,t) + \epsilon^3 g^\pm_3(x,t) +...\,,\\
h^\pm(x,t)& = &\epsilon h_1^\pm(x,t) + \epsilon^3 h^\pm_3(x,t) +...\,,\\
f^\pm(x,t) &=& 1 + \epsilon^2 f^\pm_2(x,t) + \epsilon^4 f^{\pm}_4(x,t)...\,,
\end{eqnarray*}
we get  the one dissipaton solution.

The one dissipaton solution of system (\ref{1p})-(\ref{2q}) is
\begin{eqnarray}
p^\pm (x,t) = \frac{g^\pm}{f^{\mp}} = \mu_0\frac{e^{\mu_0\eta^{\pm}_1}}{1 + b_2^{\mp} e^{\mu_0(\eta^+_1 + \eta^-_1)}}\,,\\
q^\pm (x,t) = \frac{h^\pm}{f^{\pm}} = \mu_0\frac{a^{\pm}_1  e^{\mu_0\eta^{\pm}_1}}{1 + b_2^{\pm} e^{\mu_0(\eta^+_1 + \eta^-_1)}}\,,
\end{eqnarray}
where
\begin{equation}
\eta^\pm_1 = k^\pm_1 x + \omega^\pm_1 t + \eta^\pm_{1_{0}} = \pm \left( \frac{1}{a^\pm_1} x - a^\pm_1 t           \right) + \eta^\pm_{1_0},
\end{equation}
$\omega^\pm_1 = \mp  a^\pm_1$, $k^\pm_1 = \pm 1/a^\pm_1$, 
\begin{equation}
b^+_2 = \frac{(a^+_1)^2 a^-_1 }{(a^+_1 - a^-_1)^2},\,\,\,\,\,\,\,\,
b^-_2 = \frac{(a^-_1)^2 a^+_1 }{(a^+_1 - a^-_1)^2},
\end{equation}
and
$a^\pm_1$, $\eta^\pm_{1_0}$ are real constants. Regularity of this solution requires that
$a^+_1 > 0$, $a^-_1 > 0$ and as follows $k^+_1 > 0$, $k^-_1 < 0$.

   The one dissipaton solution in laboratory coordinate system takes the form
\begin{eqnarray}
p^\pm = \left( \frac{1-v}{1+v}\right)^{\frac{1}{4}} \frac{\mu_0 k \sqrt{\sqrt{k^2 +1} \pm k}}{\cosh \left(\mu_0 k \frac{X - X_{0_\mp} -v T}{\sqrt{1-v^2}}\right)} 
e^{\pm \mu_0\left[ \frac{\sqrt{k^2 +1}}{\sqrt{1-v^2}}  (T - v X) + \nu_{1_0}  \right]} \,,
\label{1disp}\\
q^\pm = \left( \frac{1+v}{1-v}\right)^{\frac{1}{4}} \frac{\mu_0 k \sqrt{\sqrt{k^2 +1} \pm k}}{\cosh \left(\mu_0 k \frac{X - X_{0_\pm} -v T}{\sqrt{1-v^2}}\right)} 
e^{\pm \mu_0\left[ \frac{\sqrt{k^2 +1}}{\sqrt{1-v^2}}  (T - v X) + \nu_{1_0}  \right]}\,,
\label{1disq}\end{eqnarray}
where $\nu_{1_0} \equiv \eta^+_{1_0} - \eta^-_{1_0}$,
\begin{equation}
k \equiv \frac{a^+_1 - a^-_1}{2 \sqrt{a^+_1 a^-_1}}, \,\,\,\,v \equiv \frac{a^+_1 a^-_1 -1}{a^+_1 a^-_1 +1} \rightarrow |v| < 1, \label{v}
\end{equation}
and initial positions are
\begin{equation}
X_{0_\pm} = \frac{\sqrt{1-v^2}}{2k} \left(\eta^+_{1_0} + \eta^-_{1_0} + 
\ln \left(\sqrt{\frac{1+v}{1-v}}\frac{\sqrt{k^2 +1} \pm k}{4k^2}\right)\right) .
\end{equation}

In terms of parameters $v$, $k$ and $X_0 = \frac{1}{2}(X_{0_+} + X_{0_-})$, dissipaton densities become traveling wave form
\begin{eqnarray}
p^+p^- &=& \sqrt{\frac{1-v}{1+v}} \frac{2 \mu^2_0 k^2 }{\cosh\left( 2\mu_0 k \frac{X-X_0 - v T}{\sqrt{1-v^2}}\right) + \sqrt{k^2 +1}},\\
q^+q^- &=& \sqrt{\frac{1+v}{1-v}} \frac{2 \mu^2_0k^2 }{\cosh\left( 2\mu_0 k \frac{X-X_0 - v T}{\sqrt{1-v^2}}\right) + \sqrt{k^2 +1}} .
\end{eqnarray}

\subsection{Integrals of Motion}
The system (\ref{1p})-(\ref{2q}) is integrable and admits infinite number of integrals of motion. The physically meaningful are first three integrals, which can be calculated explicitly \cite{PL2022} for one dissipaton solution (\ref{1disp}),(\ref{1disq}).

The mass, momentum and energy integrals for one dissipaton solution  (\ref{1disp}),(\ref{1disq})  are
\begin{eqnarray}
M &=& \int^\infty_{-\infty} (p^+ p^- + q^+ q^-)\, dx^1 = 2 \ln  \frac{\sqrt{k^2 +1} + |k|}{\sqrt{k^2 +1} - |k|}, \\
P &= &\int^\infty_{-\infty} (p^+ \partial_1 p^- + q^+ \partial_1 q^-)\, dx^1 =\frac{4 k v}{\sqrt{1-v^2}} ,\\
E &= &\frac{4k}{\sqrt{1-v^2}}.
\end{eqnarray}

The one dissipaton solution represents relativistic particle with corresponding nonlinear mass $m_0$, momentum $P$ and energy $E$,
\begin{equation}m_0 = 4 \sinh \frac{M}{4},\,\,\,\,P = \frac{m_0 v}{\sqrt{1-v^2}},\,\,\,\,E = \frac{m_0}{\sqrt{1-v^2}},\end{equation}
and relativistic dispersion
\begin{equation}E^2 = m_0^2 + P^2  \,\,\,\rightarrow \,\,\,E = \sqrt{m_0^2 + P^2 }.
\end{equation}

\subsection{Resonant Interaction}
The above results allow us to describe
collision of two dissipatons with masses $m_1$, $m_2$ and velocities $v_1$, $ v_2$,
producing by  fusion one dissipaton with mass $m$ and velocity $v$. This process is called the resonant interaction of dissipatons.

The conservation laws for fusion of two dissipatons
$$M = M_1 + M_2,\,\,\,P = P_1 + P_2,\,\,\,\,E= E_1 + E_2, $$
are described by algebraic system of equations
\begin{eqnarray} 
\frac{\sqrt{m^2_1 + 16} + m_1}{\sqrt{m_1^2 + 16} - m_1} \,\frac{\sqrt{m^2_2 + 16} + m_2}{\sqrt{m_2^2 + 16} - m_2} &=& \frac{\sqrt{m^2 + 16} + m}{\sqrt{m^2 + 16} - m} \label{r1},\\
\frac{m_1 v_1}{\sqrt{1-v_1^2}} + \frac{m_2 v_2}{\sqrt{1-v_2^2}} &=& \frac{m v}{\sqrt{1-v^2}} \label{r2},\\
\frac{m_1}{\sqrt{1-v_1^2}} + \frac{m_2}{\sqrt{1-v_2^2}}& =& \frac{m}{\sqrt{1-v^2}}\label{r3}.
\end{eqnarray}

\subsection{Y Shaped Resonance Conditions}
We treat a special case of two equal mass $m_1 = m_2$ dissipaton collision with equal
and opposite velocities $v_1 = - v_2$.
 
The system (\ref{r1})-(\ref{r3}) has solution with $m_1 = m_2$, $v_1 = -v_2$, $v=0$
so that
\begin{equation}
v_1 = \frac{m_1}{\sqrt{m_1^2 + 16}}, \,\,\,m = \frac{1}{2} m_1 \sqrt{m_1^2 + 16}.
\label{resonance}\end{equation}

The solution describes fusion of two dissipatons with opposite momentums $P_1 = - P_2$ and the vanishing total momentum $P= P_1 + P_2 =0$. The process  creates one dissipaton with mass $m$ in the rest $v = 0$.

Due to
(\ref{v}),
the resonance condition  (\ref{resonance}) for two, one dissipaton solutions with $m_1, v_1$
and $m_2, v_2=-v_1$   restricts parameters of the solutions as
\begin{eqnarray}
a^-_1 = 1, \,\,\,a^-_2 =1, \,\,\,a^+_2 = \frac{1}{a^+_1},\,\,\,v_1 = \frac{a^+_1 -1}{a^+_1 +1}, \,\,\,v_2 = \frac{a^+_2 -1}{a^+_2 +1} = -v_1.\label{Yshapedconditions}
\end{eqnarray}

\section{Two Dissipaton Solution}
By continuing Hirota's expansion we get an exact two dissipaton solution.
  Two dissipaton solution of bilinear equations (\ref{b1})-(\ref{b4})
and the corresponding system
(\ref{1p})-(\ref{2q}) is
 \begin{eqnarray*} g^{\pm} &=& e^{\eta^\pm_1} + e^{\eta^\pm_2} + c^\pm_{112} e^{\eta^+_1 + \eta^-_1 + \eta^\pm_2} +  c^\pm_{221} e^{\eta^+_2 + \eta^-_2 + \eta^\pm_1} , \\
	 h^{\pm} &=& a^\pm_1 e^{\eta^\pm_1} + a^\pm_2 e^{\eta^\pm_2} + d^\pm_{112} e^{\eta^+_1 + \eta^-_1 + \eta^\pm_2} +  d^\pm_{221} e^{\eta^+_2 + \eta^-_2 + \eta^\pm_1},\\
f^{\pm} &=&1 + b^{\pm}_{11} e^{\eta^+_1 + \eta^-_1} + b^\pm_{12} e^{\eta^+_1 + \eta^-_2} + b^\pm_{21} e^{\eta^+_2 + \eta^-_1} + b^\pm_{22} e^{\eta^+_2 + \eta^-_2} 
+ b^{\pm}_{1122} e^{\eta^+_1 + \eta^-_1 + \eta^+_2 + \eta^-_2} 
\end{eqnarray*}
where in the light cone frame (i = 1,2)
$$  \eta^\pm_i = k^\pm_i x + \omega^\pm_i t  + \eta^\pm_{0_i}= \pm \left( \frac{1}{a^\pm_i} x - a^\pm_i t\right)   + \eta^\pm_{0_i},             $$
$\omega^\pm_i = \mp a^{\pm}_i$, $k^\pm_i = \pm 1/a^{\pm}_i$, 
and in the laboratory frame
$$  \eta^+_i + \eta^-_j = -2 k_{ij}   \frac{X - v_{ij} T}{\sqrt{1 - v^2_{ij}}} +                                          \eta^\pm_{0_i} + \eta^\pm_{0_j},$$
and 
\begin{equation}
k_{ij} = \frac{a^+_i - a^-_j}{2 \sqrt{a^+_i a^-_j}},\,\,\,\,
v_{ij} = \frac{a^+_i a^-_j -1}{a^+_i a^-_j +1}.
\end{equation}

For regularity of the solution we choose $a^+_i > 0$, $a^-_i > 0$, ($i=1,2$) and as follows, 
velocities in all frames $v_{ij}$ are bounded $|v_{ij}| < 1$. Parameters of solution are
$$c^\pm_{112}  = \frac{(a^\pm_2 - a^{\pm}_1)^2 (a^{\mp}_1)^3}{(a^\pm_2 - a^{\mp}_1)^2 (a^+_1 - a^{-}_1)^2} , 
 c^\pm_{221}  = \frac{(a^\pm_2 - a^{\pm}_1)^2 (a^{\mp}_2)^3}{(a^\mp_2 - a^{\pm}_1)^2 (a^+_2 - a^{-}_2)^2} , $$ 

$$d^\pm_{112}  = \frac{(a^\pm_2 - a^{\pm}_1)^2 (a^{\mp}_1)^2 a^\pm_1 a^\pm_2}{(a^\pm_2 - a^{\mp}_1)^2 (a^+_1 - a^{-}_1)^2} , 
 d^\pm_{221}  = \frac{(a^\pm_2 - a^{\pm}_1)^2 (a^{\mp}_2)^2 a^\pm_1 a^\pm_2}{(a^\mp_2 - a^{\pm}_1)^2 (a^+_2 - a^{-}_2)^2},  $$ 

$$ b^\pm_{ii}  =   \frac{(a^\pm_i)^2 a^\mp_i}{(a^+_i - a^-_i)^2},\,\,\,(i=j),$$  
$$     b^+_{ij}  =   \frac{(a^+_i)^2 a^-_j}{(a^+_i - a^-_j)^2}, \,\,\,    b^-_{ij}  =   \frac{(a^-_j)^2 a^+_i}{(a^+_i - a^-_j)^2},\,\, (i \neq j),        $$
$$ b^\pm_{1122} =  \frac{(a^{+}_{2}-a^{+}_1)^2 (a^{-}_{2}-a^{-}_1)^2 (a^\pm_1)^2 (a^\pm_{2})^2 a^{\mp}_1 a^{\mp}_2}{(a^{+}_{1}-a^{-}_1)^2 (a^{+}_{2}-a^{-}_2)^2 (a^{+}_{1}-a^{-}_2)^2 (a^{+}_{2}-a^{-}_1)^2} .                            $$
In above formulas we have chosen $\mu_0 =1$, which corresponds to unit "time-like" vector (\ref{velocity}). This can be accomplished by rescaling space-time and field variables:
\begin{equation}
\mu_0 x^0 \equiv x'^0,\,\,\,\mu_0 x^1 \equiv x'^1,\,\,\,\,p^\pm \equiv \mu_0 p'^\pm,\,\,\,\,q^\pm \equiv \mu_0 q'^\pm.
\end{equation}
\subsection{Y Shaped Resonance Solution}
By using two dissipaton solution, now we describe fusion of dissipatons as Y shaped resonance. 

The two dissipatonton solution under restriction on parameters 
(\ref{Yshapedconditions}), such that $c^-_{112} = c^-_{221} = d^-_{112} = d^-_{221} = b^\pm_{1122} =0 $, takes the form
\begin{eqnarray*} q^+q^- = \frac{\frac{8v^2_1}{(1-v^2_1)^2}\left[1 + \frac{1-v^2_1}{2} e^{-\frac{2 v^2_1 (T-T_0)}{1-v^2_1}} \cosh \left(\frac{2v_1 (X-X_0)}{1-v^2_1} + \frac{1}{2} \ln \frac{1-v}{1+v}\right)\right]}{\left[\cosh \frac{4 v_1 (X-X_0)}{1-v^2_1}  + \frac{1+v^2_1}{1-v^2_1}   + 
\frac{2\, e^{-\frac{2 v^2_1 (T-T_0)}{1-v^2_1}}}{\sqrt{1-v^2_1}} \cosh \frac{2v_1 (X-X_0)}{1-v^2_1}   + \frac{1}{2}  e^{-\frac{4 v^2_1 (T-T_0)}{1-v^2_1}}       \right]  }              , \end{eqnarray*}
\begin{eqnarray*} p^+p^- = \frac{\frac{8v^2_1}{(1-v^2_1)^2}\left[1 + \frac{1-v^2_1}{2} e^{-\frac{2 v^2_1 (T-T_0)}{1-v^2_1}} \cosh \left(\frac{2v_1 (X-X_0)}{1-v^2_1} - \frac{1}{2} \ln \frac{1-v}{1+v}\right)\right]}{\left[\cosh \frac{4 v_1 (X-X_0)}{1-v^2_1}  + \frac{1+v^2_1}{1-v^2_1}   + 
\frac{2\, e^{-\frac{2 v^2_1 (T-T_0)}{1-v^2_1}}}{\sqrt{1-v^2_1}} \cosh \frac{2v_1 (X-X_0)}{1-v^2_1}   + \frac{1}{2}  e^{-\frac{4 v^2_1 (T-T_0)}{1-v^2_1}}       \right]  }              . \end{eqnarray*}
This solution is called the Y shaped resonant solution.

The Y shaped resonant solution for $q^+q^-$ at $T \rightarrow -\infty$ describes collision of two dissipatons with equal and opposite velocities, the one dissipaton is moving from the left in frame $\xi^- =X - v_1 T$,
\begin{equation}
q^+ q^- = \frac{2 v^2_1}{1-v_1}
\frac{1}{\sqrt{1-v^2_1} 
\cosh ( 2v_1 \frac{X - X^L_0-v_1 T}{1-v^2_1} )
+1},
\end{equation}
and another one is moving from the right in frame $\xi^+ = X + v_1 T$,
\begin{equation}
q^+ q^- = \frac{2 v^2_1}{1+v_1}
\frac{1}{\sqrt{1-v^2_1} 
\cosh ( 2v_1 \frac{X - X^R_0 +v_1 T}{1-v^2_1}  )
+1}.
\end{equation}
After fusion, at time $T \rightarrow \infty$ we have only one dissipaton in the rest,
\begin{equation}
q^+ q^- = \frac{8 v^2_1}{1-v^2_1}
\frac{1}{(1-v^2_1)
\cosh ( 4 v_1 \frac{X - X_0}{1-v^2_1}  )
+(1 + v^2_1)}.
\end{equation}
Initial positions of dissipatons are connected by the mean value
\begin{equation}
X_0 = \frac{1}{2} (X^L_0 + X^R_0).
\end{equation}

The Y shaped resonant solution for $p^+p^-$ at $T \rightarrow -\infty$ describes collision of two dissipatons with equal and opposite velocities, the one dissipaton is moving from the left in frame $\xi^- =X - v_1 T$,
\begin{equation}
p^+ p^- = \frac{2 v^2_1}{1+v_1}
\frac{1}{\sqrt{1-v^2_1} 
\cosh ( 2v_1 \frac{X - X^L_0-v_1 T}{1-v^2_1} )
+1},
\end{equation}
and another one is moving from the right in frame $\xi^+ = X + v_1 T$,
\begin{equation}
p^+ p^- = \frac{2 v^2_1}{1- v_1}
\frac{1}{\sqrt{1-v^2_1} 
\cosh ( 2v_1 \frac{X - X^R_0 +v_1 T}{1-v^2_1}  )
+1}.
\end{equation}
After fusion, at time $T \rightarrow \infty$ we have only one dissipaton in the rest,
\begin{equation}
p^+ p^- = \frac{8 v^2_1}{1-v^2_1}
\frac{1}{(1-v^2_1)
\cosh ( 4 v_1 \frac{X - X_0}{1-v^2_1}  )
+(1 + v^2_1)}.
\end{equation}

\subsubsection{Time Reflection}

The system of equations (\ref{1p})-(\ref{2q}) is invariant under time - reflection 
\begin{equation}
T \rightarrow -T,\,\,\,\,p^\pm \rightarrow q^{\mp}, \,\,\,\,q^{\pm} \rightarrow p^\mp
\end{equation}
and as follows 
\begin{equation}
T \rightarrow -T,\,\,\,\,p^+ p^- \rightarrow q^{+}q^-, \,\,\,\,q^{+} q^- \rightarrow p^+p^-.
\end{equation}

Due to the time reflection symmetry the Y shaped solution describes also the resonant fission of one dissipaton to two dissipatons, moving in oppposite direction with equal speed.

\paragraph{Notes and Comments.}
Here we like to note intriguing difference between linear and nonlinear modes for system (\ref{1p})-(\ref{2q}). The linearized form of the equation (it could be realized as limit $\mu_0 \rightarrow \infty$) is the Klein-Gordon equation 
\begin{equation}
(-\partial^2_0 + \partial^2_1 + \mu^2_0) \Phi =0,
\end{equation}
with tachyonic dispersion $E = \sqrt{p^2 - \mu_0^2}$ and corresponding wave/particle is moving with speed bigger than characteristic speed ("speed of light"): $|v| >1 $. In contrast to this, as we have seen the nonlinear dissipaton modes, are moving with speed $|v| <1$. This property could shed light on hypothetical particles as tachyons in relativity theory. By taking nonlinear corrections to tachyons it is possible to create bradyon particles moving with speed below the speed of light.

Finally, by applying Madelung transform, dissipatons of non-relativistic  reaction-diffusion equations \cite{PL2002}  can represent resonant envelope solitons  of NLS with quantum potential. It implies that should exist another, explicit soliton resonant form for our system (\ref{1p})-(\ref{2q}) and corresponding soliton resonances. 
This work was supporting by BAP project 2022IYTE-1-0002.

%
%

\end{document}